\documentclass[manuscript]{emulateapj}

\pdfoutput=1

\newcommand\beq{\begin{equation}}
\newcommand\eeq{\end{equation}}
\newcommand\beqar{\begin{eqnarray}}
\newcommand\eeqar{\end{eqnarray}}

\shorttitle{EBL Absorption Feature in the Blazar Component of EGRB}
\shortauthors{Venters, Pavlidou \& Reyes} 
 
\begin{document}

\title{The Extragalactic Background Light Absorption Feature in the Blazar Component of the Extragalactic Gamma-ray Background}

\author{Tonia M. Venters\altaffilmark{$\ast$,1}, Vasiliki Pavlidou\altaffilmark{2,3} \& Luis C. Reyes\altaffilmark{4}}

\altaffiltext{$\ast$}{moira@uchicago.edu}
\altaffiltext{1}{Department of Astronomy and Astrophysics, The University 
of Chicago, Chicago, IL 60637}
\altaffiltext{2}{
Department of Astronomy,
The California Institute of Technology, Pasadena, CA 91125}
\altaffiltext{3}{Einstein Fellow}
\altaffiltext{4}{
Kavli Institute of Cosmological Physics and Enrico Fermi Institute, 
The University of Chicago, Chicago, IL 60637}

\begin{abstract}
High-energy photons from cosmological emitters suffer attenuation due to pair production interactions with the extragalactic background light (EBL).  The collective emission of any high-energy emitting cosmological population will exhibit an absorption feature at the highest energies.  We calculate this absorption feature in the collective emission of blazars for various models of the blazar gamma-ray luminosity function (GLF) and the EBL.  We find that models of the blazar GLF that predict higher relative contributions of high-redshift blazars to the blazar collective spectrum result in emission that is more susceptible to attenuation by the EBL, and hence result in more prominent absorption features, allowing for better differentiation amongst EBL models.  We thus demonstrate that observations of such an absorption feature will contain information regarding both the blazar GLF and the EBL, and we discuss
tests for EBL models and the blazar GLF that will become possible with upcoming \emph{Fermi} observations. 
\end{abstract}


\maketitle 

\section{Introduction}

The Energetic Gamma-ray Experiment Telescope (EGRET) aboard the \emph{Compton Gamma-ray Observatory} observed the gamma-ray sky between 1991 and 2000 at energies between 30 MeV and $\sim$ 10 GeV.  The EGRET gamma-ray sky consisted of 271 resolved gamma-ray sources included in the third EGRET Catalog of Point Sources (Hartmann et al. 1999) and the diffuse gamma-ray emission comprised of emission from the galaxy and from the extragalactic gamma-ray background (EGRB).  The origins of the EGRB are, as yet, unknown; however, since EGRET observed a number of resolved, extragalactic point sources, it is expected that unresolved sources of the same populations comprise sizable contributions to the EGRB.

Of the 271 resolved point sources observed by EGRET, 93 were identified, either confidently or potentially, as blazars (gamma-ray-loud active galactic nuclei) and in its first few months of observations, the \emph{Fermi Gamma-ray Space Telescope} has already identified 108 blazars (Abdo et al. 2009).  Thus, blazars comprise the largest class of identified gamma-ray emitters.  As such, unresolved blazars are expected to have a significant contribution to the EGRB.  The exact amount of this contribution remains undetermined due to the uncertainty in the distribution of blazars in redshift and luminosity space, the blazar gamma-ray luminosity function (GLF; Padovani et al. 1993; Stecker et al. 1993; Salamon \& Stecker 1994; Chiang et al. 1995; Stecker \& Salamon 1996, hereafter SS96; Kazanas \& Perlman 1997; Chiang \& Mukherjee 1998; Mukherjee \& Chiang 1999; M{\"u}cke \& Pohl 2000; Kneiske \& Mannheim 2005; Dermer 2007; Giommi et al. 2006; Narumoto \& Totani 2006, hereafter NT06). Hence, to this day, it is still unclear  whether the collective unresolved blazar emission comprises the bulk of the EGRB or only a small fraction of it. 

In addition to the dependence on the blazar GLF, the blazar contribution to the EGRB also depends on the distribution of blazar spectral indices at GeV energies.  The spread in the blazar spectral index distribution (SID) determines the fraction of blazars with hard spectra, which will contribute most significantly at high energies and hence will introduce curvature in the shape of the unresolved blazar emission (SS96; Venters \& Pavlidou 2007, hereafter VP07; Pavlidou \& Venters 2008, hereafter PV08). However, as with the blazar GLF, the blazar SID is also uncertain due to the low number of EGRET blazars. The uncertainty in the blazar SID results in an uncertainty in the shape of the collective unresolved blazar spectrum. Thus, it remains unclear whether blazars can simultaneously account for the high-energy emission and the low-energy emission.

The extragalactic background light (EBL) is composed of photons from starlight (at optical and ultraviolet wavelengths) and reprocessed starlight (at infrared wavelengths). At observed energies beyond the EGRET energy range, photons suffer significant attenuation due to pair production interactions with the soft photons of the EBL (Salamon \& Stecker 1998; Chen, Reyes \& Ritz 2004; Kneiske et al. 2004; Stecker et al. 2006, 2007; Franceschini et al. 2008; Primack et al. 2008; Gilmore et al. 2009). Thus, any cosmological population emitting high-energy gamma rays will exhibit an absorption feature at the highest energies in its collective spectrum.  The strength of such an absorption feature will depend on the distribution of sources with respect to redshift and luminosity. If the relative contribution of high-redshift, high-energy emitters to the collective emission is significant, the feature will be quite prominent.

Through the absorption of high-energy photons, interactions with EBL photons will produce pairs of electrons and positrons, which will Inverse Compton scatter EBL photons to high energies. The upscattered photons will, in turn, pair produce off of other EBL photons, and the process continues until the energies of the resulting photons are low enough that pair production is no longer efficient.  For the collective emission of a cosmological, gamma-ray emitting population, the effect of the ``electromagnetic cascade'' emission results in a suppression at high energies and an enhancement at lower energies. In the case of blazars, which could comprise a sizable contribution to the EGRB, predictions of the resulting enhancement at lower energies could overproduce the EGRB if the high-energy emission is high and\slash or the EBL emission is high (Coppi \& Aharonian 1997). Thus, in order to fully appreciate the blazar contribution to the EGRB, we need to include the effects of high-energy attenuation.

While the effects of the full cascade are outside the scope of this paper (see T. M. Venters 2009, in preparation), we revisit the absorption feature of the blazar cumulative emission. Specifically, we seek to determine the possible implications of the observation of the blazar absorption feature for our understanding of blazars as a cosmological population and the EBL. As previously indicated, the absorption feature depends on the GLF, the SID, and the EBL model. However, all of these inputs (including the EBL model) remain quite uncertain. Thus, we seek to demonstrate that the study of the absorption feature could be used to constrain the inputs of the collective emission and the information they can provide about blazars and the EBL.

Finally, the study of the blazar absorption feature could provide insight into the intrinsic blazar spectrum or the possible participation of multiple populations in producing the EGRB. In the era of \emph{Fermi}, improved blazar number statistics will provide much stronger constraints on GLF models. If the absorption feature in the collective unresolved blazar emission is sensitive to the GLF, then future studies of the \emph{Fermi} observations of the EGRB could have implications for blazar spectra or the relative contributions of multiple populations. For instance, if the observed absorption feature is more prominent than expected from the favored GLF, then one might suspect that blazar spectra break above some energy (e.g., spectral cutoffs, which as noted in Abdo et al. (2009), have been observed in some cases by \emph{Fermi}). On the other hand, if the observed feature is less prominent than expected, then one might suspect that other gamma-ray sources play a significant role in the production of the EGRB.

In this paper, we revisit the absorption feature of the collective unresolved blazar emission at high energies in investigating the effects of the blazar GLF and SID and the EBL model. In Section 2, we present the formalism of the calculation of the collective unresolved blazar emission. In Section 3, we discuss the inputs of the calculation and their uncertainties. In Section 4, we present the results of the calculation, and we discuss these results in Section 5.

\section{Formalism}

We define the blazar GLF per unit comoving density of blazars at GeV energies, $\rho_{\gamma}(L_{\gamma},z)$, through the following expression:
\begin{equation}\label{fe}
\rho_{\gamma}(L_{\gamma},z)p_{L}(\alpha)= \frac{d^3N}{dL_{\gamma}dV_{\rm com}d\alpha}\,,
\end{equation}
where $L_\gamma$ is the gamma-ray luminosity in ${\rm erg \, \, s^{-1}}$
at a fiducial energy $E_f$ (or, equivalently, $E_f^2$ times the
differential photon luminosity $dN_{\gamma}/dtdE$ measured at $E_f$), $V_{\rm com}$ is the comoving volume, and $p_{L}(\alpha)$ is the luminosity-independent distribution of blazar spectral indices (spectral index distribution, SID).  In order to determine the blazar luminosity-independent SID, one must correct the blazar SID for errors in measurement in gamma-ray spectral indices (VP07) and for biases introduced through determining the SID from a flux-limited sample of blazars.  The resulting $p_L(\alpha)$ is given by
\begin{equation}
p_{L}(\alpha) = \frac{\hat{p}(\alpha)}{\hat{M}(\alpha)},
\end{equation}
where $\hat{p}(\alpha)$ is the SID corrected for measurement error in the spectral indices, 
\begin{equation}
\hat{M}(\alpha) \propto \int_{F_{\gamma,\rm min}}^{\infty} dF_\gamma
\frac{1}{F_\gamma}\int_{z=0}^{\infty} dz \hat{\rho}_\gamma (\alpha,z,F_\gamma)
\frac{dV_{\rm com}}{dz}(z)
\end{equation}
 is the correction for the sample bias (see Appendix \ref{appendixa}),
\begin{eqnarray}
\hat{\rho}_\gamma  &=& L_\gamma \times \rho_\gamma(L_\gamma,z) \nonumber \\
&& \!\!\!\! \!\! = 4\pi D^2 (\alpha - 1) (1+z)^\alpha E_f F_\gamma \nonumber \\
&& \times \rho_\gamma [4\pi D^2 (\alpha - 1) (1+z)^\alpha E_f F_\gamma, z]\,,
\end{eqnarray}
and $D$ is the distance measure for the standard $\Lambda$CDM cosmology.
For an isotropic distribution of sources, the number of objects with luminosities between $L_\gamma$ and
$L_\gamma+dL_\gamma$ and spectral indices between $\alpha$ and $\alpha + d\alpha$ residing within a spherical shell at redshift $z$
with radial extent $dz$ is 
\begin{equation}
dN = \rho_\gamma (L_\gamma, z)p_L(\alpha)\,dL_\gamma\frac{dV_{\rm com}}{dz}\,dz \,d\alpha.
\end{equation}
A blazar of gamma-ray luminosity $L_\gamma$ at a redshift $z$
with a power-law source spectrum defined by the spectral index $\alpha$ 
has a photon flux of (see Appendix \ref{appendixb})
\begin{eqnarray}\label{fluxoneblazr}
F_{\rm 1,ph} (E_0, z, L_\gamma, \alpha) &=& \frac{L_\gamma}{4 \pi E_f^2
  [d_L(z)]^2}(1+z)^{2-\alpha}\left(\frac{E_0}{E_f}\right)^{-\alpha}
\nonumber \\
&& \times \exp\left[-\tau(E_0,z)\right]\!,
\end{eqnarray}
where $d_L(z)$ is the luminosity distance, defined in the concordance cosmology by
\begin{equation}
d_L(z) = \frac{c}{H_0} (1+z) \!\! \int_0^z \!\! \left[\Omega_\Lambda+\Omega_m(1+z')^3\right]^{-1/2}\,dz'\,,
\end{equation}
and $\tau(E_0,z)$ is the optical depth due to pair production on the EBL for gamma rays of observer-frame energy $E_0$ originating at redshift $z$. 
The total contribution of blazars to the gamma-ray background
{\it if we ignore secondary emission from cascades of primary gammas
  due to interactions with the EBL} can be expressed as
\begin{equation}\label{theone}
I_{E}(E_0)\! = \!\!\! \int \!\!\!\! \int \!\!\!\! \int \!\! F_{\rm ph,1}(E_0,z,L_{\gamma},\alpha)\rho_{\gamma}p_L(\alpha)\frac{d^2V_{\rm com}}{dzd\Omega}\,dL_{\gamma}\,dz\,d\alpha,
\end{equation}
where $I_{E}(E_0)$ is the intensity of the collective unresolved blazar emission given in units of photons $\mbox{s}^{-1} \mbox{cm}^{-2} \mbox{sr}^{-1} \mbox{GeV}^{-1}$.

\section{Inputs}

\subsection{Models for the Extragalactic Background Light}

The EBL intensity at the present
epoch $\left(z=0\right)$  provides an integral
constraint on the history of electromagnetic energy release in the universe since recombination. Measurement of this cumulative output however, cannot address its evolution and thus cannot be related to issues such as  the history of star and element formation. For this reason, several models have been developed to calculate the EBL luminosity density as a function of redshift, $\mathcal{L}\left(\nu,z\right)$, from fundamental astrophysical principles. These models were composed with varying degrees of complexity, observational constraints, and data inputs. 

The calculation of the high-energy absorption feature in the blazar component of the EGRB requires a model of the EBL and its evolution over cosmic time. The great degree of uncertainty associated with the EBL models and their predictions renders selecting the ``best'' candidate model impossible. Therefore, we use several models with widely ranging predictions in order to bracket the range of possible EBL realizations. 

Kneiske et al. (2004) treat the EBL-modeling problem using separate
approaches at the UV--optical and infrared wavelengths. In determining the EBL at various wavelengths, they make use of a cosmic chemical evolution model at the UV--optical wavelengths and a backwards evolution model for the infrared (see Hauser \& Dwek (2001) for a complete review of the different types of models). Additionally, this hybrid model was parameterized in terms of the main
observational uncertainties such as the redshift dependence of the cosmic
star formation rate and the fraction of UV radiation released from
star forming regions. Thus, the Kneiske et al. EBL model consists of several {\em flavors} that allow for the inclusion of various EBL scenarios that are consistent with the available data.  Specifically, the {\em Best-Fit} model best interpolates the data with the important caveat that the assumed complete UV absorption by interstellar gas introduces a sharp cutoff at $0.1\mu {\rm m}$. In the {\em Stellar-UV} model, all the UV radiation produced by the stellar populations escapes to the intergalactic medium after
reprocessing by the interstellar gas, with the {\em High-Stellar-UV} model allowing for a strong UV-field
at high redshifts.  Since the $\gamma$-ray sources likely responsible for the EGRB at GeV energies are 
particularly sensitive to the EBL density at UV wavelengths, for the purposes of this analysis, the {\em Best-Fit} and {\em High-Stellar-UV} models are used to bracket the possible ranges of attenuation.

Primack et al. (2008) have pioneered the use of semianalytical models that attempt to reproduce the process of structure formation and evolution through simulations.  Recent iterations of this model incorporate highly precise knowledge of the local luminosity
density at optical--UV (Gilmore et al. 2009) and NIR (Primack et al. 2008) wavelengths and a well-established cosmological
model. The key parameters in their approach (those that govern the rate
of star formation, supernova feedback, and metallicity) have been adjusted
to fit the local galaxy data.  With respect to estimates by Primack et al. from previous years, this
version of the model yields a lower luminosity density at optical
wavelengths, thereby resulting in a reduced EBL density. Recent
TeV observations of nearby blazars seem to support such low values (Aharonian et al. 2006).

Finally, we consider the most recent EBL model by Stecker et al. (2006). In this model, Stecker et al. calculate the EBL at infrared
and optical--UV wavelengths separately. At infrared wavelengths, they
use a backwards evolution model based on observational knowledge of:
(1) luminosity-dependent galaxy SEDs, (2) galaxy luminosity functions,
and (3) parameterized functions for luminosity evolution. 
For optical--UV wavelengths, they consider the redshift
evolution of stellar populations with an analytical approximation
to the more sophisticated SEDs used in Salamon \& Stecker (1998).
The SEDs adapted from Bruzual \& Charlot (1993) reflect stellar population synthesis
models for galaxy evolution and the observational fact that star forming
galaxies are {}``bluer'' (brighter in the blue part of the optical
spectrum) at $z>0.7$. Notably, the UV spectra for all SEDs are
assumed to cut off at the Lyman limit, and the effects of extinction
by dust are not included in the model. The former is a matter of debate
since it is not really known how much UV radiation short of the
Lyman limit can escape from the star-forming regions, while the latter
would result inexorably in an overprediction of the UV photon density
and, consequently, the optical depth at higher redshifts.  In a similar vein, Franceschini et al. (2008) also employ a backwards evolution model based rather detailed observations.  However, their determination of the EBL departs significantly from that of Stecker et al., particularly at the optical and UV wavelengths to which GeV photons are most sensitive.  The differences in these models are likely due to differences in the treatment of galaxy evolution.

EBL attenuation is a function of the observed $\gamma$-ray energy
$E_0$ and the redshift $z$ of the emitting source. The attenuation
is generally parameterized by the optical depth $\tau\left(E_0,z\right)$,
which is defined as the number of e-fold reductions of the observed
flux, $F_{\mathrm{obs}}$, as compared with the emitted source flux,
$F_{\mathrm{emitted}}$, at redshift $z$:\begin{equation}
F_{\mathrm{obs}}=e^{-\tau\left(E_0,z\right)}F_{\mathrm{emitted}}\label{eq:ebl_attenuation}\,.
\end{equation}

The optical depth is calculated from physical principles. Using the
cross section for pair production $\sigma$, and assuming isotropic
background radiation with spectral density $n\left(\epsilon\right)$
at energy $\epsilon$, the absorption probability of $\gamma$-rays
per unit path is given by \begin{equation}
\frac{d\tau}{dl}=\int_{0}^{2\pi}\sin\theta d\theta\int_{\epsilon_{\rm th}}^{\infty}n\left(\epsilon\right)\sigma\left(E_0,\epsilon,\theta\right)d\epsilon\label{eq:dtau_dl}\end{equation}
where $\theta$ is the scattering angle for the $\gamma-\gamma$ collision,
$\epsilon_{\rm th}=2m^{2}c^{4}/\left[E\left(1-\cos\theta\right)\right]$ is
the energy threshold for the reaction, and $m$ is the electron mass.
Since blazars are the sources being considered, redshift
is a good choice to measure the distance, with the total distance
being the look-back time (times the speed of light, $c$)\begin{eqnarray}
L & = & \int_{0}^{z}dz\frac{dl}{dz}\nonumber \\
 & = & \int_{0}^{z} \!\!\! dz\, \frac{c}{H_{0}\left(1+z\right)}\! \left[\left(1+z\right)^{2}\left(1+\Omega_{M}z\right)-z\left(2+z\right)\Omega_{\Lambda}\right]^{-1/2}\label{eq:dtau_dz}\nonumber \\ \end{eqnarray}
 where $H_{0}$, $\Omega_{M}$, and $\Omega_{\Lambda}$ are the well known cosmological
parameters. 

Using the expressions above, the optical depth can be written as a
function of the observed energy $E_0$ and the redshift of the emitting
source\begin{eqnarray}
\tau\left(E_0,z\right) & = & \int_{0}^{L}\frac{d\tau}{dl}dl=\int_{0}^{z}dz'\frac{dl}{dz'}\frac{d\tau\left(E',z'\right)}{dl}\label{eq:optical_depth}\\
 & = & \int_{0}^{z}\!\!\! dz'\frac{dl}{dz'}\int_{0}^{2\pi}\!\!\! \sin\theta'd\theta'\int_{\epsilon'_{\rm th}}^{\infty} \!\!\!\! d\epsilon\,\, n\left(\epsilon',z'\right)\,\sigma\left(E',\epsilon',\theta'\right)\,, \nonumber \end{eqnarray}
 where the primed variables ($E'$,$\epsilon'$ $n\left(\epsilon',z'\right)$,
$\theta'$) refer to the values calculated in the comoving frame at
$z=z'$.

\subsection{The Blazar Gamma-ray Luminosity Function}

While the blazar GLF is probably one of the most studied and debated properties of the blazar population, to this day, it remains uncertain. Model GLFs are typically constructed from luminosity functions in other wavelengths (most notably, radio or X-ray) 
exploiting a possible association of gamma-ray emission with emission in these wavelengths while correcting for possible differences in sizes of emission regions between wavelengths. However, since blazar gamma-ray emission is also, as yet, not well understood, it is unclear which lower energy luminosity function(s) could be applicable to gamma-ray--loud blazars. Typically, a model for blazar emission is assumed and, hence, a functional form of the luminosity function is adopted. For instance, if one assumes the synchrotron self-Compton model for blazar gamma-ray emission, then one would expect that the low-energy synchrotron emission in blazars would be closely related to the gamma-ray emission. The unknown parameters (e.g., normalization due to relativistic beaming, the faint-end slope of the luminosity function) are subsequently fitted to gamma-ray data (see e.g., SS96; NT06; Giommi et al. 2006). 

While such a procedure represents the best that can be done with current data, it should be noted that a great degree of uncertainty remains as several issues remain unresolved. Since it is more difficult to observe fainter objects (which are crucial in the calculation of the collective unresolved blazar calculation), any flux-limited sample will be biased toward brighter objects. Thus, uncertainties will always be larger in the faint-end slope of the luminosity function. Additionally, there is some indication that not all blazars are necessarily explained by the same emission process and that different types of blazars (i.e., BL Lacertae-like objects and flat-spectrum radio quasars) could form separate populations with respect to emission (Sikora et al. 2002; B\"{o}ttcher 2007) and, hence, require separate luminosity functions (M{\"u}cke \& Pohl 2000; Dermer 2007). There is also the possibility that flaring blazars (and different \emph{types} of flaring blazars) could also form separate emission populations with respect to emission and require separate luminosity functions (SS96). Finally, due to the large positional error circles, there are many unidentified EGRET sources which could also have a sizable contribution to the EGRB (Pavlidou et al. 2008). A number of these unidentified sources could, in fact, be (and are fairly likely to be) \emph{unidentified blazars}. Thus, one might underestimate the normalization of gamma-ray blazars with respect to low-energy blazars in not accounting for the possibility of not being able to identify some resolved blazars; in addition, the existence of resolved but unidentified blazars would introduce uncertainties to the redshift distribution of resolved blazars, which is one of the constraints that luminosity functions are typically required to satisfy.

With the availability of \emph{Fermi} data, many more blazars will be observed, and at least some of the aforementioned uncertainty will be alleviated. However, for now, with these caveats in mind, for the purposes of this paper, we perform our calculations using the best-guess pure luminosity evolution (PLE) and the luminosity-dependent density evolution (LDDE) models of NT06. It should be noted that the functional form of the PLE model of the blazar GLF originated from radio data, while the functional form of the LDDE model originated from X-ray data. Thus, if the calculated blazar absorption feature is sensitive to the blazar GLF, then the observation of such a feature could be used as an additional constraint to the preferred luminosity functional form and, by extension, to blazar emission models.

\subsection{The Spectral Index Distribution}

The unresolved blazar contribution to the EGRB is not just a question of magnitude, but also of spectral shape, and the spectral shape is sensitive to the distribution of blazar spectral indices at GeV energies.  If all blazars had the same spectral index then the spectrum of unresolved emission would be simply a power law.  If, on the other hand, the SID has some spread, the spectrum will have some curvature (SS96; PV08).  If the spread is small, then even if the blazar contribution dominates the EGRB at lower energies, it may not be enough to explain the emission at higher energies (PV08).  Thus, in order to answer the question of the unresolved blazar contribution to the EGRB, the blazar SID has to be determined and the resulting shape calculated.

Obtaining the SID of blazars is complicated by the presence of large errors in measurement of individual blazar spectral indices.  If these errors are not properly taken into account, sampling of the underlying \emph{intrinsic} spectral index distribution (ISID) will be contaminated by exaggerating its spread, leading, in turn, to exaggeration in the curvature of the unresolved collective emission (VP07).  Furthermore, the presence of spectrally distinct populations of blazars (e.g,. flaring versus quiescent, BL Lac objects versus flat-spectrum radio quasars) can also contaminate the ISID.  In order to determine the ISID of the collective blazar population while minimizing the contamination due to measurement errors, VP07 performed a  likelihood analysis fitting the third EGRET data set of confident blazars to a Gaussian ISID.  They determined that the maximum-likelihood Gaussian ISID can be characterized by a mean ($\alpha_0$) of 2.27 and a spread ($\sigma_0$) of 0.2.  They also performed the analysis by dividing the sample of confident blazars into their subpopulations, flaring versus quiescent and BL Lac objects versus FSRQs.  In the case of flaring and quiescent blazars, they found no evidence that the subpopulations are spectrally distinct (though the lack of adequate time resolution made dividing the subpopulations difficult).  In the case of BL Lac objects and FSRQs, they found a marginal $1 \sigma$ separation between BL Lac objects ($\alpha_0 = 2.15$, $\sigma_0 = 0.28$) and FSRQs ($\alpha_0 = 2.3$, $\sigma_0 = 0.19$).  They also found that the flaring and quiescent blazar populations are spectrally consistent.

In PV08, the shapes of the unresolved emission were calculated for the collective blazar population and BL Lac objects and FSRQs.  In the cases of the collective population and FSRQs, the curvatures of the shapes were not enough to allow the populations to explain all of the EGRB, though in the case of BL Lac objects, the curvature was enough to, in principle, allow BL Lac objects to explain the EGRB.  However, in all cases, the normalizations of the emission were not determined, and the uncertainties in the shapes resulting from the uncertainties in the likelihood analysis are considerable.

It should be noted that since the PLE and LDDE GLF models do not distinguish between the subpopulations of blazars, for the purposes of self-consistency, we also do not  distinguish between them with regards to their SIDs.  Thus, for the purposes of this analysis, we include only the collective blazar population ISID of VP07(correcting for biases introduced in sampling a flux-limited catalog).  Notably, \emph{Fermi} has already provided evidence that BL Lac objects and FSRQs are spectrally distinct (Abdo et al. 2009). However, since the \emph{Fermi} blazar catalog is not yet complete, it is currently premature to construct luminosity functions (especially those that distinguish between BL Lac objects and FSRQs) from \emph{Fermi} data. In recognizing the importance of correctly treating spectral distinctions among subpopulations of blazars, we will return to this issue in a future publication.

\section{Results} 

The blazar contributions to the EGRB as calculated assuming two separate models of the blazar GLF and several models of the EBL are plotted in Figure 1. The black lines represent contributions determined assuming the LDDE model of the blazar GLF, while the gray lines represent contributions determined assuming the PLE model of the blazar GLF. For comparison, the blazar contributions assuming no absorption (solid lines) and the Strong et al. (2004) determination of the EGRET EGRB (data points with statistical error bars) are also plotted. Since the GLFs used include the maximum-likelihood parameters determined by NT06, the blazar contributions comprise $\sim 50$\% of the overall background\footnote{However, as discussed in Section 3.2, the determinations of the GLFs and their parameters remain highly uncertain. In the likelihood analysis performed by NT06, the most likely level of unresolved emission from blazars is $\sim 25\%-50$\% of the EGRB.  Nevertheless, in several of the cases presented in NT06, parameters for which unresolved blazars can account for $100$\% of the background are within the $1\sigma$ contours.}. As demonstrated in SS96 and PV08, there is considerable curvature in the unresolved blazar emission due to the spread in the blazar SID indicating the increasing importance of blazars with harder spectral indices at higher energies\footnote{There is some uncertainty in the determination of the parameters of the blazar spectral index, which will result in uncertainty in the overall shape of the spectrum of the unresolved blazar emission.  As indicated in PV08, this uncertainty in the spectral shape is quite large for EGRET blazars, but will improve considerably with \emph{Fermi} observations. For this reason, and because \emph{Fermi} is currently taking data, we simply calculate the EBL absorption for the best guess spectra.}. 

\begin{figure}[t]
\begin{center}
\resizebox{3.25in}{!}{\includegraphics{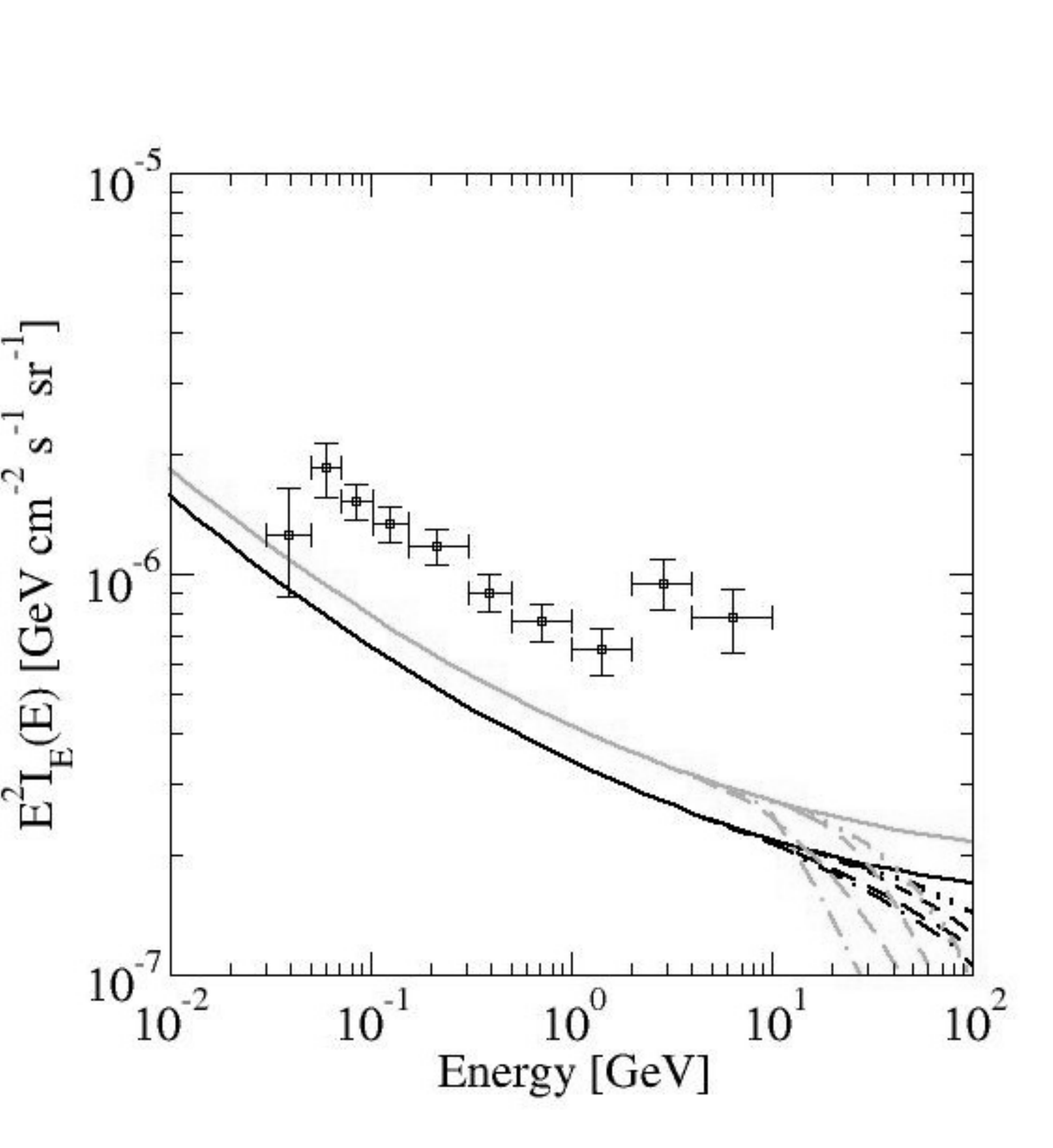}}
\vspace{0.1in}
\caption{\label{fig1} Collective gamma-ray emission of unresolved blazars as a function of observed energy calculated for PLE and LDDE models of the blazar GLFs and several models of the EBL.  \emph{Black:} the collective emission for blazars assuming the LDDE GLF. \emph{Gray:} the collective emission assuming the PLE GLF.  \emph{Solid:} the collective emission assuming no absorption. \emph{Thin dashed:} the collective emission including absorption assuming the Kneiske et al. (2004) best-fit EBL model. \emph{Heavy dashed:} the collective emission including absorption assuming the Kneiske et al. (2004) EBL model with the high UV component. \emph{Dotted:} the collective emission including absorption assuming the Gilmore et al. (2009) EBL model. \emph{Dot dashed:} the collective emission including absorption assuming the Stecker et al. (2006) EBL model. \emph{Double dot-dashed:} the collective emission including absorption assuming the Franceschini et al. (2008) EBL model.}
\end{center}
\end{figure}

The most striking feature in Figure 1 is that of the suppression at high energies due to the considerable amount of absorption by pair production interactions with EBL photons. Certain EBL models are quite distinguishable from the others.  Most notably, the Kneiske et al. (2004) high UV model and the Stecker et al. (2006) model predict a greater degree of absorption than the other three models. This is due to the fact that the Kneiske high UV model and the Stecker model predict a higher amount of UV background radiation than the others.  Since the pair production cross section as a function of the  center-of-mass energy peaks at the electron mass, one would expect that gamma-ray photons of energy $\sim$ tens of GeV are most likely to interact with UV background photons. Thus, unsurprisingly, models with high UV backgrounds will result in more suppression at high energies.

Another striking observation from Figure 1 is that the high-energy suppressions for the PLE model are consistently steeper than those of the LDDE model. The different appearances of the features can be explained by considering that the blazar GLF is the distribution of blazars in luminosity and redshift space. Since the PLE model suppressions are steeper than those of the LDDE model, one would conclude that high-redshift blazars contribute more to the high-energy emission in the PLE model than in the LDDE model. In investigating this possibility, we plot the unresolved emission evaluated at several energies as a function of redshift (Figure 2). As in Figure 1, the black lines represent the emission assuming the LDDE model of the blazar GLF, and the gray lines represent the emission assuming the PLE model of the blazar GLF. As can be seen in Figure 2, the emission for the LDDE model peaks at lower redshifts ($z \sim 0.05$) than that for the PLE model ($z \sim 0.6$)\footnote{In both cases, the redshift where the emission peaks increases as the minimum luminosity included in the integration increases. This is to be expected since high-luminosity objects that contribute to the EGRB will be distributed toward higher redshifts. The inclusion of low-luminosity blazars allows more low-redshift objects to participate.}. Thus, high-energy photons in the PLE model suffer more attenuation due to interactions with the EBL photons than in the LDDE model. Additionally, the emission is more sharply peaked in the LDDE model than in the PLE model indicating that the participating blazars in the LDDE model are more concentrated to a particular epoch while those in the PLE model are more spread out over epoch.

\begin{figure}[t]
\begin{center}
\resizebox{3.25in}{!}{\includegraphics{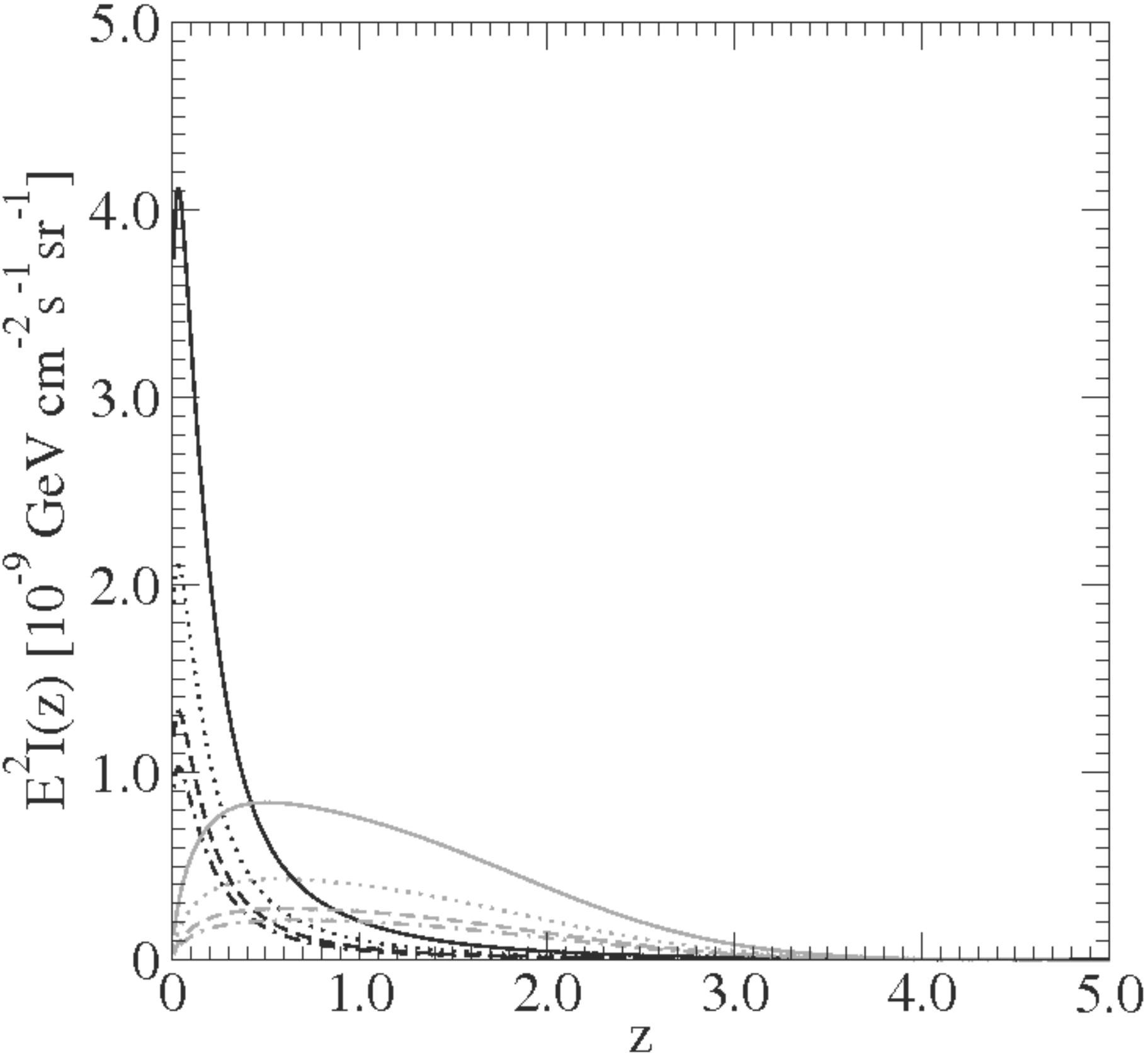}}
\vspace{0.1in}
\caption{\label{fig2} Collective gamma-ray emission of unresolved blazars as a function of redshift calculated for the PLE and LDDE models of the blazar GLF and evaluated at several energies (not including absorption). \emph{Black:} the collective emission for blazars assuming the LDDE GLF. \emph{Gray:} the collective emission assuming the PLE GLF.  \emph{Solid:} the collective emission as a function of \emph{z} evaluated at an observed energy of $100$ MeV. \emph{Dotted:} the collective emission as a function of \emph{z} evaluated at an observed energy of $1$ GeV. \emph{Dashed:} the collective emission as a function of \emph{z} evaluated at an observed energy of $10$ GeV. \emph{Dot dashed:} the collective emission as a function of \emph{z} evaluated at an observed energy of $100$ GeV.}
\end{center}
\end{figure}

\section{Discussion and Conclusions}

We have calculated the blazar contribution to the EGRB including attenuation of high-energy photons due to interactions with the EBL. We have found that (a) the EBL attenuation of high-energy gamma rays results in a suppression in the spectrum of the unresolved emission; (b) the shape of the high-energy suppression depends on the blazar GLF model and the EBL model; and (c) the high-energy suppression for the PLE model is steeper than that for the LDDE model indicating that higher-redshift blazars contribute more to the emission at high energies in the PLE model than in the LDDE model.

As demonstrated in this paper, the suppression at high energies can be a probe for the underlying redshift--luminosity distribution of gamma-ray blazars. If, in fact, blazars do comprise the bulk of the EGRB, then this suppression should be observed in the EGRB that will be measured by \emph{Fermi}. Thus, the observation of the EBL absorption feature at high energies could provide information about the relative contribution of blazars to the EGRB.  

However, it is conceivable that any features observed in the EGRB could also be attributable to other effects.  Multi-wavelength observations of blazars reveal the characteristic doubly-peaked synchrotron/Compton structures in blazar spectra.  Thus, while it is true that when viewed at small energy intervals in the GeV band blazar spectra appear as single power laws, we do not expect this observation to remain true as the high-energy end of the observational energy interval increases.  Instead, blazar spectra are expected (and, as noted in Abdo et al. (2009), in certain cases have already been observed by {\it Fermi}) to break beyond GeV energies. Such spectral breaks will also manifest as a feature in the collective emission of blazars.  In this case, the feature would be sensitive to the nature of blazar spectra at high energies, which, in turn, encode information about blazar emission mechanisms.  There is also the possibility that blazars could account for the bulk of the emission at lower energies, but not at higher energies. If the collective gamma-ray emission of another gamma-ray source \emph{peaks} at high energies, then we would also expect to observe a high-energy feature due to the transition to a different population.  An intriguing example of this scenario would include high-energy emission from dark matter annihilation, which peaks near the mass of the dark matter particle (e.g., Ando et al. 2007; Siegal-Gaskins \& Pavlidou 2009).  In order to distinguish between the EBL suppression feature and features arising from other possible scenarios, it will be essential to constrain (to the extent possible) the expected shape and strength of the high-energy absorption feature in the blazar EGRB contribution, as well as to quantify the associated theoretical uncertainties. 

Due to its substantially increased sensitivity with respect to that of EGRET, \emph{Fermi} will resolve more than an order of magnitude more blazars than EGRET did. The number of blazars that \emph{Fermi} will observe will also provide insight into the blazar GLF (see, e.g., SS96, NT06). The measured GLF will allow us to not only determine the shape of the blazar absorption feature, but also the collective blazar emission as a function of redshift. With the \emph{Fermi}-measured blazar GLF, we can compare the redshift-dependent gamma-ray emission with that of blazar GLFs determined from lower-energy observations. In so doing, we can determine how closely the gamma-ray emission in blazars is tied with the emission at lower energies. 

\emph{Fermi} measurements of blazar spectral indices would also allow further investigation into the possible existence of a spectral subpopulation of blazars with harder spectral indices such as high-frequency-peaked BL Lac objects (HBLs).  There is some speculation that the harder spectral indices of the HBLs could be indicative of a greater contribution due to a particular emission mechanism relative to the contribution due to the emission mechanism that is more prevalent for blazars with softer spectral indices\footnote{For instance, in the leptonic emission scenarios, the harder spectral indices of the HBLs could possibly indicate the greater importance of external Compton emission relative to the synchrotron self-Compton emission that could dominate the high-energy emission in blazars with softer spectral indices, such as the FSRQs.}. So far, HBLs have mainly been observed at low redshifts, and as such, one would expect the absorption feature of the collective spectrum to be not very prominent. However, if \emph{Fermi} measurements reveal that there are more hard blazars at high redshifts relative to soft blazars than indicated by EGRET data, then the effect of their intrinsic spectra would be to flatten the overall collective spectra of blazars. Moreover, the collective spectrum of their subpopulation would exhibit quite a prominent absorption feature, and hence, their high-energy emission would further impact the collective blazar spectrum through electromagnetic cascade radiation (T. M. Venters 2009, in preparation). Furthermore, investigating the absorption features of distinct spectral subpopulations of blazars could provide intriguing insight into blazar gamma-ray emission and their evolution with cosmic time.

Thus, the study of the collective unresolved blazar emission from \emph{Fermi} observations should provide important constraints about both blazars and other possible sources of gamma-ray emission.

In this paper, we computed only the attenuation of the high-energy emission due to interactions with the EBL. However, when high-energy photons interact with EBL photons, they initiate electromagnetic cascades, which generate emission at lower energies. However, an accurate inclusion of the such secondary emission requires detailed Monte-Carlo simulations of the EBL-induced cascades and is outside the scope of this paper. We plan to return to this problem in a future publication (T. M. Venters 2009, in preparation). Furthermore, while the \emph{Fermi Gamma-ray Space Telescope} has begun taking data and has already observed many more blazars than EGRET did, a complete catalog of blazars has not yet been released by the Fermi Collaboration.  Thus, the determination of the blazar GLF(s) from \emph{Fermi} data is, as yet, premature.  With this in mind, rather than making a solid prediction for the anticipated \emph{Fermi} results, we have demonstrated the sensitivity of the shape of the spectrum suppression due to EBL absorption to various model inputs, thus making explicit the information content of such a feature. 

\acknowledgements{We acknowledge enlightening feedback from Angela Olinto, Joel Primack, and Floyd Stecker.  This work was supported in part by the Kavli Institute for
  Cosmological Physics at the University of Chicago through grants NSF
  PHY-0114422 and NSF PHY-0551142 and an endowment from the Kavli
  Foundation and its founder Fred Kavli. T.M.V. was supported by an
  NSF Graduate Research Fellowship.  V.P. acknowledges support by NASA through the \emph{GLAST} Fellowship Program, NASA Cooperative Agreement: NNG06DO90A.}

\appendix

\section{A.  Sample Bias Correction $\hat{M}(\alpha)$}\label{appendixa}

The ISID $\hat{p}(\alpha)$ obtained by VP07 using EGRET data is measured using a roughly flux-limited data set (to the extent that we can postulate that EGRET resolved all blazars of integrated gamma-ray flux greater than some value $F_{\gamma, \rm min}$ and none with smaller $F_\gamma$), so that 
\begin{equation}\label{a1}
\hat{p}(\alpha) = \frac{1}{N_{\rm tot}} 
\int_{F_\gamma=F_{\gamma,\rm min}}^\infty \frac{d^2N}{dF_\gamma d\alpha} dF_\gamma\,
\end{equation}
where $N_{\rm tot}$ is the total number of objects in the sample.
However, this is not the SID $p_L(\alpha)$ that enters Eq. \ref{theone}. The latter is defined by Eq. (\ref{fe}) and is the distribution of spectral indices for blazars in a luminosity interval between $L_\gamma$ and $L_\gamma+dL_\gamma$. A flux-limited sample will be biased toward harder spectral indices than a fixed gamma-ray luminosity interval, because not all blazars with the same $L_\gamma$ have the same flux: harder blazars have higher fluxes in the high-energy band and are more easy to detect.  A relation can be derived between $\hat{p}(\alpha)$ and $p_L(\alpha)$, starting from a relation between $d^2N/dF_\gamma d\alpha$ and $d^3N/dL_\gamma dV_{\rm com} d\alpha$:
\begin{equation}\label{step1}
\frac{d^2N}{dF_\gamma d\alpha} = \int_{z=0}^{\infty} dz\frac{d^3N}{dL_\gamma dV_{\rm com}d\alpha}\left|\frac{\partial(L_\gamma,V_{\rm com},\alpha)}{\partial(F_\gamma,z,\alpha)}\right|\,.
\end{equation}
$L_\gamma$ is proportional to $F_\gamma$ multiplied by a function of $z$ and $\alpha$ (see Eq. \ref{lfrelation}). In addition, in writing Eq. (\ref{fe}), we have assumed that $\alpha$ is independent of $L_\gamma$ and $z$. Thus, we obtain
\begin{equation}\label{step2}
\left|\frac{\partial(L_\gamma,V_{\rm com},\alpha)}{\partial(F_\gamma,z,\alpha)}\right| = 
\left|\frac{\partial L_\gamma}{\partial F_\gamma}\frac{dV_{\rm com}}{dz}\right| = 
\frac{L_\gamma}{F_\gamma}\left|\frac{dV_{\rm com}}{dz}\right|\,,
\end{equation}
since
\begin{equation}
L_{\gamma} = 4\pi d_L^2 (\alpha-1)(1+z)^{\alpha-2} E_f F_{\gamma}\,.
\end{equation}
Eq. (\ref{step2}) combined with Eqs. (\ref{step1}) and (\ref{fe}), gives
\begin{equation}
\frac{d^2N}{dF_\gamma d\alpha} = p_L(\alpha) \frac{1}{F_\gamma}\int_{z=0}^{\infty} dz L_\gamma \rho_{\gamma}(L_\gamma,z)\frac{dV_{\rm com}}{dz}\,.
\end{equation}
Substituting into Eq. (\ref{a1}) we obtain 
\begin{equation}
\hat{p}(\alpha) = \frac{1}{N_{\rm tot}}p_L(\alpha)
\int_{F_{\gamma,\rm min}}^{\infty} dF_\gamma
\frac{1}{F_\gamma}\int_{z=0}^{\infty} dz L_\gamma \rho_{\gamma}(L_\gamma,z)\frac{dV_{\rm com}}{dz} \equiv
p_L(\alpha) \hat{M}(\alpha)
\end{equation}
where the last equality gives the definition of the sample correction bias $\hat{M}(\alpha)$; the normalization, $N_{\rm tot}$, is obtained by requiring that $p_L(\alpha)$ integrates to 1. 

\section{B. Derivation of the Blazar Contribution to the Extragalactic Gamma-ray Background}\label{appendixb}

The isotropic gamma-ray luminosity of a blazar at some fiducial rest-frame energy, $E_f$ (the energy emitted in photons of energy $E_f$ per unit time, assuming that the blazar emits isotropically), is related to its integrated photon flux, $F_\gamma$ (the number of photons emitted in energies above {\em observer} frame energy $E_f$ per unit time per unit area), through 
\begin{equation}\label{lfrelation}
L_{\gamma} = 4\pi d_L^2 (\alpha-1)(1+z)^{\alpha-2} E_f F_{\gamma},
\end{equation}
where $d_L$ is the luminosity distance, and we have assumed that the 
blazar has a single--power-law energy spectrum ($dN_{\gamma}/dE_{\gamma} \propto E_{\gamma}^{-\alpha}$).
  In turn, the differential single-blazar flux, $F_{\rm ph,1}(E_0)$ (number of photons per unit energy per unit time per unit area emitted at observer-frame energy, $E_0$), is related to $F_\gamma$ through 
\begin{equation}
F_{\gamma} = \int_{E_f}^{\infty} F_{\rm ph,1}(E_0)\,dE_0.
\end{equation}
Having assumed a power-law spectrum, $F_{\gamma}$ becomes
\begin{equation}
F_{\gamma} = \int_{E_f}^{\infty} F_{\rm ph,1}(E_f) \left(\frac{E_0}{E_f} \right)^{-\alpha}\,dE_0 = \frac{E_f F_{\rm ph,1}(E_f)}{\alpha - 1}\,.
\end{equation}
Substituting the above into the equation for $L_{\gamma}$ and solving for $F_{\rm ph,1}(E_f)$, we get
\begin{equation}
F_{\rm ph,1}(E_f) = \frac{L_{\gamma}}{4\pi d_L^2 E_f^2}(1+z)^{2-\alpha}.
\end{equation}
Thus, neglecting absorption, $F_{\rm ph,1}(E_0)$ is given by
\begin{equation}
F_{\rm ph,1}(E_0) = F_{\rm ph,1}(E_f)\left(\frac{E_0}{E_f}\right)^{-\alpha} = \frac{L_{\gamma}}{4\pi d_L^2 E_f^2}(1+z)^{2-\alpha}\left(\frac{E_0}{E_f}\right)^{-\alpha}.
\end{equation}
Including the effects of absorption, we arrive at Eq. (\ref{fluxoneblazr}) of Section 2:
\begin{equation}
F_{\rm ph,1}(E_0) =  \frac{L_{\gamma}}{4\pi d_L^2 E_f^2}(1+z)^{2-\alpha}\left(\frac{E_0}{E_f}\right)^{-\alpha} e^{-\tau (E_0,z)}.
\end{equation}
The intensity of emission is defined as
\begin{equation}
I_E(E) = \frac{d^4N_{\gamma}}{dtdAdEd\Omega} = \frac{d}{d\Omega} \int \!\! dN F_{\rm ph,1},
\end{equation}
where $N_{\gamma}$ is the number of photons, $N$ is the number of objects, and $F_{\rm ph,1}$ is the flux from a single contributing object.
Making the dependencies explicit, the intensity can be expressed as
\begin{equation}
I_E(E_0) = \frac{d}{d\Omega} \int F_{\rm ph,1}(E_0,z,L_{\gamma},\alpha) \frac{d^3 N}{dL_{\gamma}  dV_{\rm com} d\alpha}\,dL_{\gamma}\,dV_{\rm com}\,d\alpha.\
\end{equation}
The differential number of objects can be expressed in terms of the GLF and the ISID:
\begin{equation}
\frac{d^3N}{dL_{\gamma}dV_{\rm com} d\alpha} = \rho_{\gamma}(L_{\gamma},z)p_L(\alpha).
\end{equation}
The comoving volume element is given by (assuming $\Lambda$CDM cosmology)
\begin{equation}
\frac{d^2V_{\rm com}}{dzd\Omega} = \frac{c}{H_0}D^2[\Omega_{\Lambda}+\Omega_m(1+z)^3]^{-1/2},
\end{equation}
where $D = d_L/(1+z)$ is the distance measure.
Substituting for $F_{\rm ph,1}$, $dV_{\rm com}$, and the differential number of objects, we finally arrive at the sought for expression for the emission {\em without attenuation}:
\begin{equation}
I_E(E_0) = \frac{c}{H_0} \frac{1}{4\pi E_f^2} \!\! \int \! d\alpha\,p_L(\alpha)\left(\frac{E_0}{E_f}\right)^{-\alpha}\!\!\! \int \! dz\,(1+z)^{-\alpha}[\Omega_{\Lambda}+\Omega_m(1+z)^3]^{-1/2} \!\! \int \! dL_{\gamma}L_{\gamma}\rho_{\gamma}.
\end{equation} The above expression for the collective unresolved blazar emission is equivalent to Equation (10) of SS96, corrected for cosmology and assuming that blazars consist of a single spectral population.

Including the attenuation factor, $\exp\left[-\tau(E_0,z)\right]$, in the expression for the single-blazar flux, we obtain:
\begin{equation}
I_E(E_0) = \frac{c}{H_0} \frac{1}{4\pi E_f^2} \!\! \int \! d\alpha\,p_L(\alpha)\left(\frac{E_0}{E_f}\right)^{-\alpha}\!\!\! \int \! dz\,e^{-\tau(E_0,z)}(1+z)^{-\alpha}[\Omega_{\Lambda}+\Omega_m(1+z)^3]^{-1/2} \!\! \int \! dL_{\gamma}L_{\gamma}\rho_{\gamma}.
\end{equation}


\begin{thebibliography}{}

\bibitem[Fermi LAT Collaboration (2009)]{fer09} Abdo, A. et al. (Fermi LAT Collaboration)\ 2009, \apj, 700, 597

\bibitem[Ando et al.(2007)]{2007PhRvD..75f3519A} Ando, S., Komatsu, E., 
Narumoto, T., \& Totani, T.\ 2007, \prd, 75, 063519 

\bibitem[B\"{o}ttcher (2007)]{bot07} B\"{o}ttcher, M.\ 2007, \apss, 309, 95

\bibitem[Bruzual \& Charlot (1993)]{bru93} Bruzual, A.~G. \& Charlot, S.\ 1993, \apj, 405, 538

\bibitem[Chiang et al.(1995)]{chi95} Chiang, J., Fichtel, C.~E., von Montigny, C., Nolan, P.~L., \& Petrosian, V.\ 1995, \apj, 452, 156

\bibitem[Chiang \& Mukherjee(1998)]{chi98} Chiang, J., \& Mukherjee, R. 1998, Astrophys. J., 496, 752

\bibitem[Chen et al.(2004)]{2004ApJ...608..686C} Chen, A., Reyes, L.~C., 
\& Ritz, S.\ 2004, \apj, 608, 686 

\bibitem[Coppi \& Aharonian(1997)]{cop97} Coppi, P. \& Aharonian, F.~A.\ 1997, \apj, 487, L9  

\bibitem[Dermer (2007)]{der07} Dermer, C.~D.\ 2007, \apj, 659, 958

\bibitem[Franceschini et al. (2008)]{fra08} Franceschini, A., Rodighiero, G., \& Vaccari, M.\ 2008, A\&A, 487, 837

\bibitem[Giommi et al. (2006)]{gio06} Giommi, P., Colafrancesco, S., Cavazzuti, E., Perri, M., \& Pittori, C.\ 2006, A\&A, 445, 843

\bibitem[Gilmore et al. (2009)]{gil09} Gilmore, R.~C., Madau, P., Primack, J.~R., Somerville, R.~S., \& Haardt, F.\ 2009, arXiv:0905.1144

\bibitem[Hartmann et al.(1999)]{har99} Hartmann, R.~C., et al.\ 
1999, \apjs, 123, 79 

\bibitem[Hauser \& Dwek (2001)]{hau01} Hauser, M.~G. \& Dwek, E.\ 2001, ARA\&A, 39, 249

\bibitem[Kazanas \& Perlman (1997)]{kaz97} Kazanas, D., \& Perlman, E.\ 1997, \apj, 476, 7

\bibitem[Kneiske et al. (2004)]{kne04} Kneiske, T.~M., Bretz, T., Mannheim, K., \& Hartmann, D.~H.\ 2004, A\&A, 413, 807

\bibitem[Kneiske \& Mannheim (2005)]{KM05} Kneiske, T.~M., Mannheim, K.\ 2005,
  Proceedings of the 29th International Cosmic Ray Conference, 4, 1

\bibitem[M{\"u}cke \& Pohl(2000)]{mp00} M{\"u}cke, A., \& 
Pohl, M.\ 2000, \mnras, 312, 177 

\bibitem[Mukherjee \& Chiang(1999)]{mc99} Mukherjee, R., \& 
Chiang, J.\ 1999, Astroparticle Physics, 11, 213 

\bibitem[Narumoto \& Totani(2006)]{NT06} Narumoto, T., \& 
Totani, T.\ 2006, \apj, 643, 81 

\bibitem[Padovani et al.(1993)]{pad93} Padovani, P., 
Ghisellini, G., Fabian, A.~C., \& Celotti, A.\ 1993, \mnras, 260, L21 

\bibitem[Pavlidou et al.(2007)]{petal07} Pavlidou, V., Siegal-Gaskins,
  J.~M, Fields, B.~D., Olinto, A.~V., \& Brown, C.\ 2008, \apj, 677, 27

\bibitem[Pavlidou \& Venters(2008)]{pv08} Pavlidou, V. \& Venters, T.~M.\ 2008, \apj, 673, 114

\bibitem[Primack et al. (2008)]{pri08} Primack, J.~R., Gilmore, R.~C., \& Somerville, R.~S.\ 2008, arXiv:0811.3230

\bibitem[Salamon \& Stecker(1994)]{sal94} Salamon, M.~H., \& 
Stecker, F.~W.\ 1994, \apjl, 430, L21

\bibitem[Salamon \& Stecker(1998)]{sal98} Salamon, M.~H., \& 
Stecker, F.~W.\ 1998, \apj, 493, 547 

\bibitem[Siegal-Gaskins 
\& Pavlidou(2009)]{sgp09} Siegal-Gaskins, J.~M., \& Pavlidou, V.\ 2009, \prl, 102, 241301

\bibitem[Sikora et al.(2002)]{2002ApJ...577...78S} Sikora, M., 
B{\l}a{\.z}ejowski, M., Moderski, R., 
\& Madejski, G.~M.\ 2002, \apj, 577, 78 

\bibitem[Stecker et al. (1993)]{ste93} Stecker, F.~W., Salamon, M.~H., \& Malkan, M.~A.\ 1993, \apjl, 410, L71

\bibitem[Stecker \& Salamon(1996)]{ste96} Stecker, F.~W., \& 
Salamon, M.~H.\ 1996, \apj, 464, 600 

\bibitem[Stecker et al. (2006)]{ste06} Stecker, F.~W., Malkan, M.~A., \& Scully, S.~T.\ 2006, \apj, 648, 774

\bibitem[Stecker et al. (2007)]{ste07} ---.\ 2007, \apjs, 658, 1392

\bibitem[Strong et al.(2004)]{smr04} Strong, A.~W., 
Moskalenko, I.~V., \& Reimer, O.\ 2004, \apj, 613, 956 

\bibitem[Venters \& Pavlidou (2007)]{vp07} Venters, T.~M. \& Pavlidou,
  V.\ 2007, ApJ, 666, 128
  
\bibitem[Venters (2009)]{ven09} Venters, T.~M.\ 2009, {\it in preparation}

\end{thebibliography}
\end{document}